\documentclass[3p,times, review]{elsarticle}
\usepackage{times,graphicx}
\usepackage{soul}
\usepackage{booktabs}
\usepackage{xr}
\usepackage[hidelinks]{hyperref}

\usepackage{float}
\usepackage{amsmath}
\usepackage{caption}
\usepackage{subcaption}
\usepackage{comment}
\usepackage{upgreek}

\usepackage{cleveref}
\usepackage{amssymb}
\biboptions{comma,square,sort&compress}
\usepackage[figuresright]{rotating}

\usepackage{mathtools}
\usepackage{xcolor}
\DeclareCaptionFormat{subfig}{\figurename~#1#2#3}
\DeclareCaptionSubType*{figure}
\newcommand{\nopar}{{\parfillskip=0pt \parskip=0pt \par}}

\begin{document}
\begin{frontmatter}

\title{Atomistically informed phase field study of austenite grain growth}
\author[1]{Ayush Suhane\corref{correspondingauthor}}
\cortext[correspondingauthor]{Corresponding author}
\ead{ayush.suhane@ubc.ca}
\author[2]{Daniel Scheiber}
\author[2]{Vsevolod I. Razumovskiy}
\author[1]{Matthias Militzer}
\address[1]{Centre for Metallurgical Process Engineering, The University of British Columbia, Vancouver, BC, Canada V6T 1Z4}
\address[2]{Materials Center Leoben Forschung GmbH, Roseggerstrasse 12, Leoben 8700, Austria}

\begin{abstract}

Atomistically-informed phase field simulations have been performed  
%In this study, we utilize atomistic scale information of solute segregation in FCC-Fe to parameterize phase field simulations 
to investigate the effect of five common alloying elements (Nb, Ti, Mo, V, Mn) on austenite grain growth. The anisotropic simulations based on the segregation energy profiles of the solutes to four different grain boundary (GB) types from density functional theory calculations suggest a secondary role of solute drag anisotropy on grain growth. Hence, the solute trends are determined to be the same for all investigated GBs, and as a result, the $\Sigma 5(310)[001]$ GB can be considered as a representative GB for solute trend predictions. The decrease in grain growth rates due to solute additions is quantitatively described using a solute trend parameter. 
The following hierarchy of the solute's effectiveness to retard austenite grain growth has been determined based on the results of the presented model calculations in agreement with the experimental observations: Nb$>$Ti$>$Mo$>$V$\approx$Mn.
%The solute trends are determined as Nb$>$Ti$>$Mo$>$V$\approx$Mn in decreasing order of their effectiveness to retard austenite grain growth in agreement with the experimental observations. 
The limitations and the strengths of the proposed approach are discussed in detail, and a potential application of this approach to steel design is proposed.
%The limitations and the strengths of the proposed approach are discussed in light of current experimental knowledge. 
\end{abstract}


\end{frontmatter}
%\linenumbers

\section{Introduction}
The ever-increasing 
%societal 
demand for high-performance steel products in the transportation, energy, and construction sectors requires continuous innovation in steel design and processing strategies. As a design paradigm, Integrated Computational Materials Engineering (ICME)~\cite{council_integrated_2008} utilizes knowledge-based process models that link the processing parameters to the material properties by controlling the microstructure. During thermo-mechanical controlled processing, the microstructure is primarily modified due to grain growth, recrystallization, and phase transformations, and as a result, the development of accurate process models for these metallurgical processes is of crucial importance. Historically, time-consuming and expensive experiments have been used to parameterize empirical fit parameters with limited predictive capabilities for different steel compositions and processing conditions. Computational techniques, on the other hand, have progressed rapidly in recent years and can facilitate modeling across different length and time scales such that microstructure evolution models can be combined with atomistic simulations to develop integrated process models. Such developments may be useful for alloy design, where various alloying strategies are explored to obtain the desired microstructures with improved properties.

An important aspect of microstructure evolution is grain growth where a network of high-angle grain boundaries (GBs) evolve to decrease the grain boundary energy per unit volume of the system. The GB migration rates are extremely sensitive to the alloy content. The alloying elements and/or impurities, even in trace quantities, can either form precipitates that pin the grain boundaries or remain in solution with associated potential GB segregation, resulting in slower GB migration rates due to particle pinning~\cite{manohar_five_1998, maalekian_situ_2012} and solute drag~\cite{sinclair_effect_2007, zurob_model_2001}, respectively. The solute solubility increases with temperature, and the retardation effect due to the latter is more prominent in controlling the average grain size at higher temperatures, e.g., for Nb in low-alloyed austenite during reheating. Empirical grain growth model approaches
%, e.g., $\overline{D}_G = Kt^n$,  where $\overline{D}_G$ is the average grain diameter, $K$ is a rate parameter, and $n$ is the grain growth exponent, 
for different steel compositions have been proposed over the past few decades~\cite{uhm_prediction_2004, lee_prediction_2008}, but they have been eventually found to suffer from limited predictive capabilities~\cite{zhang_growth_2011, pous-romero_austenite_2013}. Alternatively, physical models that explicitly consider Zener pinning pressure~\cite{zener_theory_1949} due to precipitates and/or solute drag pressure due to alloying elements in solution, e.g., using the Cahn-L\"{u}cke-St\"{u}we (CLS) solute drag model~\cite{cahn_impurity-drag_1962, lucke_quantitative_1957, lucke_theory_1971}, are applicable to a relatively wide range of processing conditions and material chemistries. Sinclair et al.~\cite{sinclair_effect_2007} considered the CLS solute drag model to quantitatively describe the retardation of ferrite grain growth in Fe-0.095 wt.\% Nb in comparison to pure Fe between 700-900~$^\circ$C. Fu et al.~\cite{fu_austenite_2011} described austenite grain growth rates in Fe-0.09 wt.\% C with 0.049 and 0.09 wt.\% Nb between 950-1300 $^\circ$C using a model that considers the combined effect of precipitation pinning and solute drag. Using a similar model, Furumai et al.~\cite{furumai_evaluating_2019} systematically varied the Nb concentration between 0 to 0.06 wt.\% and quantified the decrease in grain growth rates in Fe-0.09 wt.\% C-1 wt.\% Mn between 1100-1400 $^\circ$C. In these studies, two key parameters, i.e., representative segregation energy and trans-GB diffusivity, are considered as fit parameters that describe the role of Nb in solution on migrating grain boundaries.

Atomistic calculations such as Density Functional Theory (DFT) that quantify the solute-GB interactions can be instrumental in quantifying these parameters~\cite{jin_study_2014, scheiber_ab_2015}. In our previous study~\cite{suhane_solute_2022}, we presented an approach to calculate the effective segregation energy from the binding energy profiles of selected solutes in Au from DFT calculations. The effective segregation energies are in close agreement with those determined from the CLS model to describe the migration of a single well-characterized GB in Au. Further, the effective segregation energies of several solutes, such as P, C, V, and Al in BCC-Fe determined from DFT calculations, are in reasonable agreement with the experimentally determined segregation energies~\cite{scheiber_impact_2021}.

Solute segregation also varies with the GB structure. However, DFT calculations, due to their high computational cost, are restricted to high coincident grain boundaries (low-$\mathrm{\Sigma}$ GBs). For example, Ito and Sawada~\cite{ito_first-principles_2022} determined the binding energy of transition elements (Ti, V, Cr, Mn, Co, Ni, Cu, Nb, and Mo) for nine [001] symmetric tilt GBs in FCC-Fe with increasing $\mathrm{\Sigma}$ values from 5 to 41, which is the largest data-set available in the literature for austenite. These GBs constitute only a small subset of boundaries that may be found in a polycrystalline material. Such a limitation can be mitigated by considering the success of the structural unit model where a high-$\Sigma$ GB can be geometrically described as a combination of the structural units of high-symmetric or low-$\Sigma$ GBs~\cite{sutton_structure_1983}. Dingreville et al.~\cite{dingreville_interaction_2016} demonstrated that low coincident [001] symmetric tilt GBs in Ni could be constructed using combinations of four primary structural units found in low-$\mathrm{\Sigma}$ GBs, i.e., $\mathrm{\Sigma}$5(210), $\mathrm{\Sigma}$5(310), $\mathrm{\Sigma}$13(510), and $\mathrm{\Sigma}$13(320) and determined the structure dependence of hydrogen segregation from the superposition of segregation in these structural units. As a result, the qualitative segregation behavior of general grain boundaries can, in a first approximation, be considered similar to the constituent low-$\mathrm{\Sigma}$ grain boundaries.

Grain boundaries with different binding energy profiles will migrate differently during grain growth depending on the presence of solutes due to anisotropy in solute drag. In this regard, mesoscale modeling techniques such as Monte Carlo~\cite{holm_misorientation_2001, gruber_misorientation_2009}, Phase field method~\cite{kazaryan_grain_2002, miyoshi_multi-phase-field_2017, miyoshi_large-scale_2021} and Level-set approaches~\cite{elsey_simulations_2013, fausty_2d_2020} have been used to investigate the role of anisotropic GB properties such as GB mobility, GB segregation, and GB energy on grain growth. In particular, we demonstrated with a parametric 2D phase field study that a representative segregation energy can be introduced to describe the mean grain size evolution during normal grain growth with moderate solute drag anisotropy~\cite{suhane_representative_2023}. In these cases, the solute trends for grain growth in specific alloy systems can be deduced using mesoscale simulations, provided information is available on representative segregation energies of different solutes.

As a result, the objective of the present study is to investigate the role of structure-dependent GB segregation energies, determined from DFT calculations, on the grain growth kinetics in austenite and to verify whether a representative GB can be defined to determine the solute trends. To this end, we consider the binding energy data reported by Ito and Sawada~\cite{ito_first-principles_2022} for nine symmetric tilt GB in FCC-Fe and five common alloying elements (Nb, Ti, Mo, V, Mn) in steels and perform two-dimensional phase field simulations to simulate anisotropic grain growth. The next section outlines the methodology to incorporate the atomistic binding energies of solutes in the phase field simulations. The effect of segregation anisotropy and bulk solute concentration on the austenite grain growth are presented in the following section. A representative GB is defined, and the solute trends due to the solute drag effect are verified with experimental observations. 

\section{Simulation methodology}
\subsection{Solute drag parameters}
The phenomenological approach of Cahn~\cite{cahn_impurity-drag_1962}, L\"{u}cke and St\"{u}we~\cite{lucke_quantitative_1957, lucke_theory_1971} describes the GB migration rates ($v$) in the presence of solutes and under a driving pressure using two parameters, i.e., effective segregation energy ($E$) and trans-GB diffusivity ($D$). In this approach, the solute drag pressure ($\Delta G_{SD}$) that opposes the driving pressure is given as:
\begin{equation}
    \Delta G_{SD} = \frac{\alpha c_0 v}{1 + \beta^2 v^2}
    \label{eqn:cahnsd}
\end{equation}
where, $c_0$ is the bulk solute concentration, while $\alpha$ and $\beta$ are solute drag coefficients such that,
\begin{equation}
    \alpha = \frac{2{\color{red}{\delta_{r}}} N_v (k_{\scriptstyle B}T)^2}{D E}\left[sinh\left(\frac{E}{k_{\scriptstyle B}T}\right) - \frac{E}{k_{\scriptstyle B}T}\right] 
     \quad \quad \quad \beta^2  = \frac{{\color{red}{\delta_{r}}} \alpha k_{\scriptstyle B}T}{2N_v D E^2}.
\end{equation}
Here, $N_v$ is the number of atoms per unit volume at the grain boundary, {\color{red}{2$\delta_{r}$ is the physical GB width}}, and $k_{\scriptstyle B}$ is the Boltzmann constant. Such a formulation leads to low-velocity and high-velocity branches where solutes either follow the sufficiently slow-moving GB or the GB breaks away from the solute atmosphere at high velocities. Furthermore, the solute drag pressure in this approach is independent of the sign of the effective segregation energy, i.e., both segregating and anti-segregating solutes will lead to identical solute drag pressure. It has been suggested that the effective segregation energy ($E$) can be determined from the solute enrichment (GB excess) at the stationary ($v$ = 0) GB~\cite{maruyama_interaction_2003}, that can be obtained from either experimental measurements~\cite{gupta_role_2020} or atomistic simulations~\cite{suhane_solute_2022, scheiber_ab_2015}. {\color{red}{From DFT calculations, the total GB excess ($\varGamma_{DFT}$) can be determined from the solute binding energy profile across the grain boundary as~\cite{suhane_solute_2022}}}
\begin{equation}
\frac{c^l_{\scriptstyle GB}}{1-c^l_{\scriptstyle GB}} = \frac{c_{\scriptstyle 0}}{1-c_{\scriptstyle 0}} \exp\left(\frac{-E^l_{seg}}{k_{\scriptstyle B}T}\right) \quad \quad  \quad \varGamma_{DFT} = \frac{1}{A}\sum\limits_{l}^{N_{\scriptstyle GB}} (c^l_{\scriptstyle GB}-c_{\scriptstyle 0})
\label{eqn:DFT_gamma}
\end{equation}
where $c^l_{\scriptstyle GB}$ is the solute concentration at GB site $l$ at temperature $T$ (in~K), {\color{red}{$E^l_{seg}$ is the binding energy of a solute at GB site $l$, $A$ is the grain boundary area for the unit cell, and $N_{\scriptstyle GB}$ is the number of grain boundary sites in a unit cell.}} As proposed in our previous study~\cite{suhane_solute_2022}, $E$ corresponding to the CLS approach can be determined by considering equal solute enrichment in the atomistic and the continuum model, i.e., $\varGamma_{DFT}$ = $\varGamma_{CLS}$, where the GB excess in the CLS model is given by,
\begin{equation}
    \varGamma_{CLS} = N_v \int_{-{\color{red}{\delta_{r}}}}^{{\color{red}{\delta_{r}}}} (c(x)-c_{\scriptstyle 0}) dx = \varGamma_0 c_{\scriptstyle 0} \left[\frac{k_{\scriptstyle B}T}{E}\left\{1- \exp\left(-\frac{E}{k_{\scriptstyle B}T}\right)\right\} - 1\right]
    \label{eqn:CLS_gamma}
\end{equation}
Here, $\varGamma_0 = N_v (2{\color{red}{\delta_{r}}})$ is the total number of GB sites projected onto the habit plane that is equivalent to the surface density in the continuum model. The maximum surface density from atomistic calculations can be determined from $N_{GB}/A$ that varies between 40 - 45~atoms/nm$^2$ for all the grain boundaries investigated by Ito and Sawada~\cite{ito_first-principles_2022}. Considering {\color{red}{$2\delta_{r}$}} as 0.5~nm in accordance with DFT calculations~\cite{ito_first-principles_2022}, $N_v$ is determined to vary between 81 - 90~atoms/nm$^3$ whereas $N_v$ = 90~atoms/nm$^3$ is determined from the lattice constant of FCC-Fe at 0~K. A $10\%$ variation in $N_v$ corresponds to less than 30~meV difference in effective segregation energies for all investigated grain boundaries that is smaller than the accuracy of atomistic calculations. Therefore, $N_v$ = 90~atoms/nm$^3$ is used in all the subsequent simulations. In the following discussion, the total GB enrichment ($\varGamma_{GB}$) is equivalent to both $\varGamma_{DFT}$ and $\varGamma_{CLS}$, respectively.

Relatively little is known about the trans-GB diffusivity of solutes. Previous investigations of solute effects on grain growth have suggested that the trans-GB diffusivity lies more closely to the bulk diffusivity of solute than the GB diffusivity~\cite{sinclair_effect_2007, buken_model_2017, furumai_evaluating_2019}. For instance, Sinclair et al.~\cite{sinclair_effect_2007} suggested that the trans-GB diffusion of Nb is a factor of 15 higher than its bulk diffusion in ferrite, and similarly, other researchers have suggested that the trans-GB diffusion of Nb and V is twice the bulk diffusivity in austenite~\cite{buken_model_2017, furumai_evaluating_2019}. On the other hand, GB diffusion of solutes, e.g. Ni, in both ferrite and austenite is reported to be approximately four orders of magnitude larger than the bulk diffusivity~\cite{geise_lattice_1985, hanatate_grain_1978}. Further, the volume self-diffusion of Fe in austenite is found to be approximately five orders of magnitude smaller than the grain boundary diffusivity~\cite{oikawa_lattice_1982, hanatate_grain_1978}. As a result, for trend prediction (and simplicity), we consider trans-GB diffusivity to be equivalent to the bulk diffusivity ($D_b$) of solutes in austenite as reported by {\color{red}{Kurokawa et al.~\cite{kurokawa_diffusion_1983} and}}, Oikawa~\cite{oikawa_lattice_1982}. The corresponding pre-exponential factor, $D_0$, and activation energy, $Q$, {\color{red}{and the solute diffusivity at 1273 K ($D_{1273K}$)}} is given in Table~\ref{tab:diff}.

\begin{table}[h!]
\centering
\noindent
\caption{Diffusion coefficients for the solute elements considered in the present study.}
\label{tab:diff}  
\begin{tabular}{|c|c|c|c|c|}\hline
\multicolumn{1}{|c|}{{Element}} & \multicolumn{1}{c|}{{$D_0$ (cm$^2$/s)}} & \multicolumn{1}{c|}{{$Q$ (eV)}} & \multicolumn{1}{c|}{$D_{1273 K}$ ($\mathrm{\times 10^{-12} cm^2/s}$)} & \multicolumn{1}{c|}{Ref.} \\  
\hline
\multicolumn{1}{|c|}{Nb} & \multicolumn{1}{c|}{0.75} & \multicolumn{1}{c|}{2.74} & \multicolumn{1}{c|}{10.80} & \multicolumn{1}{c|}{\cite{kurokawa_diffusion_1983}}\\
\multicolumn{1}{|c|}{Ti} & \multicolumn{1}{c|}{0.15} & \multicolumn{1}{c|}{2.61} & \multicolumn{1}{c|}{6.83} & \multicolumn{1}{c|}{\cite{oikawa_lattice_1982}}\\
\multicolumn{1}{|c|}{Mo} & \multicolumn{1}{c|}{0.036} & \multicolumn{1}{c|}{2.48} & \multicolumn{1}{c|}{5.36} & \multicolumn{1}{c|}{\cite{oikawa_lattice_1982}}\\
\multicolumn{1}{|c|}{V} & \multicolumn{1}{c|}{0.28} & \multicolumn{1}{c|}{2.73} & \multicolumn{1}{c|}{4.26} & \multicolumn{1}{c|}{\cite{oikawa_lattice_1982}} \\
\multicolumn{1}{|c|}{Mn} & \multicolumn{1}{c|}{0.18} & \multicolumn{1}{c|}{2.74} & \multicolumn{1}{c|}{2.56} & \multicolumn{1}{c|}{\cite{oikawa_lattice_1982}} \\
\hline
%\bottomrule
\end{tabular}
\end{table}

\subsection{Phase field method}
Phase field is a diffuse interface approach where each grain $i$ in a polycrystalline structure is characterized by a phase field parameter ($\eta_i$) which is a continuous function of space and time such that it takes a value of 1 inside the grain, and gradually decreases at the grain boundary to 0 outside the grain. {\color{red}{Two phase field frameworks are most common in simulating grain growth, (1) the multi-phase field model proposed by Steinbach and coworkers~\cite{steinbach_generalized_1999, steinbach_phase-field_2009}, and (2) the continuum phase field model proposed by Chen and Yang~\cite{chen_computer_1994-1}, and Fan and Chen~\cite{fan_computer_1997}. While the former model has advantages that the material properties can be directly included in the time-evolution equations of the phase field parameter, adding anisotropy in GB properties, such as grain boundary energy, results in numerical artifacts, as shown in detail by Garcke et al.~\cite{garcke_multiphase_1999}. As a result, we utilize the continuum phase field model which was further extended by Shahandeh et al.~\cite{shahandeh_friction_2012} to simulate the solute effects on grain growth in a microcrystalline system where the grain sizes are of the order of few $\upmu$m. In this model, a friction pressure is used to retard the migration rates of grain boundaries in agreement with the solute drag.}} The derivation of the model is given elsewhere~\cite{shahandeh_friction_2012}, whereas the evolution of the grain structure with friction pressure is determined using the Allen-Cahn equation as, 
\begin{equation}
    \frac{\partial \eta_i}{\partial t} = L\left[\kappa \nabla ^2 \eta_i - m\left(\eta_i^3 -\eta_i + 3\eta_i \sum_{j\neq i}^p \eta_j^2\right) - 3\eta_i\sum_{j\neq i}^p\eta_j\Delta G_{ij}\right]
    \label{eqn:pfm}
\end{equation}
where $\kappa$, $m$, and $L$ are the phase field model parameters related to GB energy ($\gamma$), GB mobility ($M$), and artificial GB width (2$\delta$) through the following equations,
\begin{equation}
\kappa = \frac{3\gamma\delta}{2} \quad \quad \quad m = \frac{3\gamma}{\delta} \quad \quad \quad L = \frac{2M}{3\delta}   
\end{equation}
and $p$ is the number of co-existing grains at any point in space, $\Delta G_{ij}$ is the friction pressure opposing the migration of the grain boundary between grain $i$ and grain $j$ that is considered equivalent to the solute drag pressure given by Eqn.~\eqref{eqn:cahnsd}. {\color{red}{Similar to}} Shahandeh et al.~\cite{shahandeh_friction_2012}, Eqn.~\eqref{eqn:pfm} is solved iteratively since the friction pressure is dependent on the GB velocity $v$ that is determined from $\partial\eta_i/\partial t$ as $v$ = $\frac{\partial \eta_i}{\partial t}/|\nabla \eta_i|$. As a result, at each grid point, $\partial\eta_i/\partial t$ from the previous timestep is used to determine the $\partial\eta_i/\partial t$ at the current time-step using Eqn.~\eqref{eqn:pfm} and Eqn.~\eqref{eqn:cahnsd}. Subsequently, a new velocity is computed, and the process is repeated until convergence in GB velocity is achieved. 

Note that the GB properties such as GB mobility ($M$), effective segregation energy ($E$), and GB energy ($\gamma$) can vary with the GB structure, however, such a detailed dependence is not available for all the grain boundaries in a realistic microstructure. A pragmatic structure dependence, which considers the GB properties to be dependent on the disorientation angle, has been considered in several studies~\cite{miyoshi_multi-phase-field_2017, liu_phase_2019, chang_effect_2014} to assess the effect of variability in GB properties on grain growth kinetics. As shown in our previous study~\cite{suhane_representative_2023}, a moderate variability in GB mobility and energy, on the one hand, and those for solute drag, on the other hand, can be separated to identify the properties of a representative GB. Thus, only the effective segregation energy ($E$) is considered anisotropic in the present study, and other GB properties, such as GB mobility and GB energy, are assumed to be isotropic for all the cases. The effective segregation energy of a solute at a grain boundary between grain $i$ and $j$ is determined as $E_{ij}$ based on the DFT data for different grain boundaries, which is then used for the friction pressure calculation using Eqn.~\eqref{eqn:cahnsd} and Eqn.~\eqref{eqn:pfm}. 

\subsection{Computational details}
An initial two-dimensional grain structure with $\sim$18000 grains is constructed using Voronoi tesselation considering periodic boundary conditions in all directions where the simulation domain of 2~mm $\times$ 2~mm is divided into a 2000~$\times$~2000 grid. As a result, the grid size ($\Delta x$) is fixed as 1~$\upmu$m. The initial microstructure is allowed to evolve considering ideal grain growth with isotropic grain boundary properties and in the absence of solute drag until it reaches the scaling grain size distribution. The scaled microstructure is considered as the starting microstructure that contains $\sim$14000 grains with an average grain size (radius) of 10~$\upmu$m. {\color{red}{To determine the grain boundary resolution, a shrinking circle is simulated as a benchmark. The shrinkage rates in simulations with four points in the grain boundary resulted in approximately 2\% deviation in comparison to the analytical solution which is much smaller than the accuracy of experimental grain size measurements, i.e., $\sim$10\%~\cite{pous-romero_austenite_2013}. Thus, the grain boundary is resolved in four grid points, and consequently, the grain boundary width (2$\delta$) is 4~$\upmu$m such that the initial average grain diameter is five times the grain boundary width.}} The representative GB energy ($\gamma_{rep}$) is taken as 1~$\mathrm{J/m^2}$ for austenite, and the representative GB mobility is assumed as $M_{rep}$~=~$1.5\times 10^{-11}$~$\mathrm{m^4/Js}$ as suggested by Furumai et al.~\cite{furumai_evaluating_2019} for $T$~=~1273~K. An isothermal heat treatment with a holding time ($t_{sim}$) of 400~s at 1273~K is considered to assess the grain growth kinetics. For all the simulations, the time step is fixed to 0.0065~s which satisfies the stability condition, i.e., $\Delta t < \Delta x^2/M_{rep}\gamma_{rep}$. The time step sensitivity is verified by considering a smaller time step, i.e., $\Delta t$ = 0.0032~s, that results in identical grain growth kinetics for ideal grain growth. In the absence of any solutes, a final grain size of 51~$\upmu$m is obtained using the above parameters in 2D phase field simulations.
\begin{figure}[t!]
    \centering
    \includegraphics[width=\textwidth, keepaspectratio]{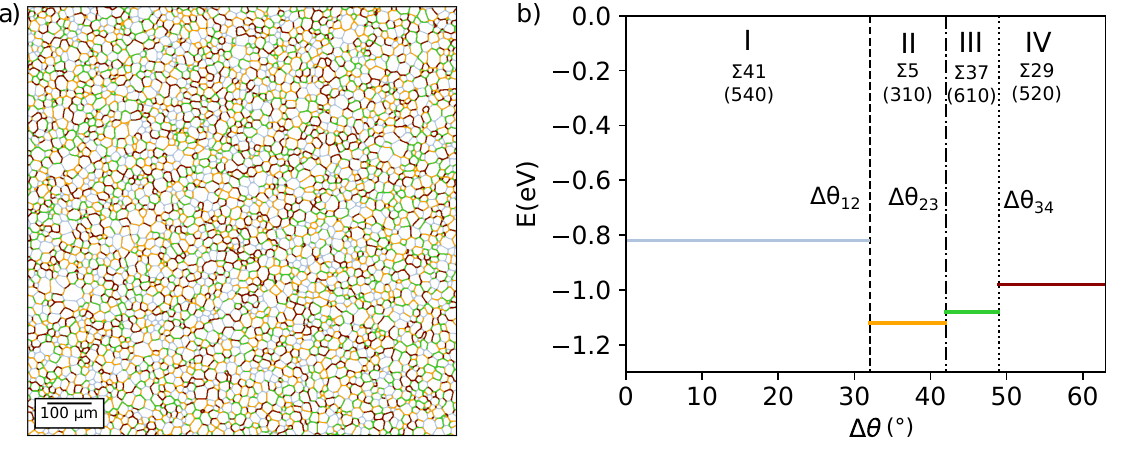}
        %\phantomsubcaption
    \vspace{-7mm}
    \caption{(a) Polycrystalline microstructure with an average grain size of 10~$\upmu$m in a domain size of 2~mm $\times$ 2~mm. Only a quarter of the simulation domain is shown for clarity. (b) Four distinct GB types, shown with different colors, are considered where each GB type has a different effective segregation energy. The effective segregation energies for ten ppm addition of Nb for different [001] tilt grain boundaries are shown as an example.}
    \label{fig:Initial}
\end{figure}
Each grain in the scaled microstructure is assigned three random Euler angles, and the disorientation angle between each pair of grains, i.e., grain $i$ and grain $j$, is determined and stored in a look-up table~\cite{miyoshi_multi-phase-field_2017}. This type of distribution leads to a texture that resembles the well-known Mackenzie distribution in disorientation for a statistically sufficient number of grains in the microstructure~\cite{liu_phase_2019}. Disorientation-dependent segregation energies are distributed in the microstructure by considering four types of grain boundaries, i.e., GB-I, GB-II, GB-III, and GB-IV, as shown in Fig.~\ref{fig:Initial}. Three cutoff disorientation angles, $\Delta \theta_{12}$, $\Delta \theta_{23}$, and $\Delta \theta_{34}$ are defined that control the fraction of a particular GB type in the microstructure due to the Mackenzie-like disorientation distribution~\cite{liu_phase_2019}, {\color{red}{e.g., $\Delta \theta_{12}$~\textless ~$\Delta \theta$~\textless~$\Delta \theta_{23}$ corresponds to GB type II}}. Fig.~\ref{fig:Initial} shows the initial microstructure that contains an equal fraction of four GB types by defining $\Delta \theta_{12}$ = 32$^\circ$, $\Delta \theta_{23}$ = 42$^\circ$, and $\Delta \theta_{34}$ = 49$^\circ$. The effective segregation energies shown in Fig.~\ref{fig:Initial} are determined for 10~ppm (atom fraction) addition of Nb in $\Sigma41(540)[001]$, $\Sigma5(310)[001]$, $\Sigma37(610)[001]$, $\Sigma29(520)[001]$ GBs, respectively.

\subsection{Phenomenological grain growth model}
Phase field simulations and, consequently, the average grain size evolution considering anisotropic segregation energy is rationalized using a mean-field grain growth model. For curvature-driven grain growth, the grain growth rate, i.e., the change in average grain size ($\overline{R}$) with respect to time, is described as the product of the representative GB mobility and the driving pressure, which can be written as,
\begin{equation}
    \frac{\partial \overline{R}}{\partial t} = k_1 M_{rep} \left( \frac{k_2 \gamma_{rep}}{\overline{R}} - k_3 \Delta G_{SD, rep}\right)
    \label{eqn:ggeq}
\end{equation}
where $k_1$, $k_2$ and $k_3$ are geometrical constants, and $\Delta G_{SD, rep}$ is the solute drag pressure due to the representative segregation energy ($E_{rep}$) in the presence of an impurity and/or alloying element. A more detailed analysis of the geometrical constants can be found in our previous study~\cite{suhane_representative_2023}. Accordingly, $k_1$ and $k_2$ are taken as 0.89 and 0.20, respectively, for all cases. For the average grain size of 10~$\upmu$m, the driving pressure is sufficiently low such that the grain boundaries either remain unaffected by solutes for relatively low solute concentrations in the bulk or are in the low-velocity limit for higher bulk concentrations. As a result, $k_3$ is taken here as 1 for all cases~\cite{suhane_representative_2023}.

\section{Results}
\subsection{Anisotropy in effective segregation energy}
\begin{figure}[H]
    \centering
    \includegraphics[width=0.5\hsize]{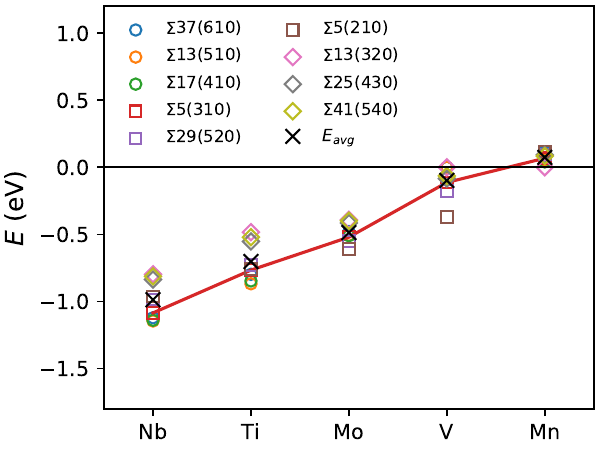}
    \caption{Compilation of effective segregation energy for different [001] symmetric tilt grain boundaries determined from DFT calculations~\cite{ito_first-principles_2022} for 10~ppm solute addition at $T$=1273 K. $E_{avg}$ corresponds to the average for all the nine grain boundaries.}
    \label{fig:e0_compile}
\end{figure}

The effective segregation energies for different [001] symmetric tilt GBs, characterized by their $\Sigma$ value, are determined from Eqn.~\eqref{eqn:DFT_gamma} and Eqn.~\eqref{eqn:CLS_gamma} using the binding energy data from DFT~\cite{ito_first-principles_2022} for bulk solute concentration of 10~ppm at 1273~K. These energies are compiled in Fig.~\ref{fig:e0_compile} for Nb, Mo, Ti, V, and Mn ordered according to 
%in decreasing order of 
their atomic sizes
~\cite{pauling_atomic_1947}. Here, a negative effective segregation energy corresponds to solute segregation, whereas a positive value corresponds to solute depletion at the grain boundary. The data indicates favorable segregation of all solutes, except Mn, in all the investigated GBs, with the largest solute, Nb, having the largest magnitude of the segregation energy. There is an appreciable variability in effective segregation energies among different GBs, e.g. the absolute difference between the maximum and minimum values is 0.34~eV, 0.38~eV, 0.21~eV, 0.37~eV, and 0.15~eV for Nb, Ti, Mo, V, and Mn, respectively. However, identical solute trends are determined for individual grain boundaries. As a reference, the effective segregation energies for the above order of solutes are determined as -1.11~eV, -0.78~eV, -0.53~eV, -0.12~eV, and 0.07~eV for the $\Sigma 5(310)[001]$ GB, and the average segregation energies ($E_{avg}$) for all the nine GBs are -1.0~eV, -0.70~eV, -0.48~eV, -0.09~eV and 0.07~eV, respectively.    

\subsection{Phase field simulations}
Two configurations of phase field simulations are considered, i.e. (1) isotropic segregation of solutes considering $\Sigma 41(540)[001]$, $\Sigma 5(310)[001]$, and $\Sigma 17(410)[001]$ as the representative GB for weak, moderate and strong segregation tendencies, respectively, and (2) anisotropic segregation by considering an equal proportion of four GBs, i.e., $\Sigma 5$(310)[001], $\Sigma 29$(520)[001], $\Sigma 37(610)[001]$ and $\Sigma 41(540)[001]$ in the initial microstructure as shown in Fig.~\ref{fig:Initial}. For the latter case, the weighted mean segregation energy of the microstructure is defined as $E_{avg, m}$ = $\int_{l_{gb}} E \delta l/\int_{l_{gb}} \delta l$, where $l_{gb}$ is the total grain boundary length, $\delta l$ refers to a grain boundary segment and $E$ to the effective segregation energy of the grain boundary segment. For these grain boundaries, $E_{avg, m}$ is within 10~meV of the average segregation energy of all nine grain boundaries and is therefore considered to be representative for anisotropic simulations.      

\begin{figure}[h]
    \centering
    \begin{subfigure}{.49\textwidth}
        \centering 
        \includegraphics[width=\hsize]{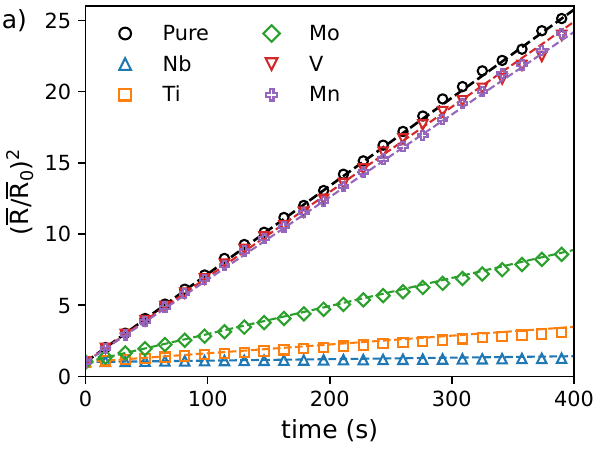}
        \phantomsubcaption
        \label{fig:gg_aniso}
    \end{subfigure}%
    \hfill
    \begin{subfigure}{.49\textwidth}
        \centering 
        \includegraphics[width=\hsize]{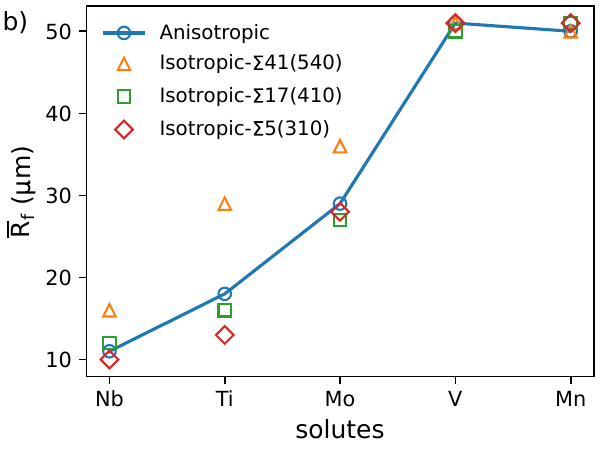}
        \phantomsubcaption
        \label{fig:finalgrainsizes}
    \end{subfigure}%
    \vspace{-7mm}
    \caption{(a) Square of normalized average grain size as a function of time as obtained from phase field simulations considering anisotropic segregation energy with an equal fraction of $\Sigma5(310)[001]$, $\Sigma29(520)[001]$, $\Sigma37(610)[001]$ and $\Sigma41(540)[001]$ GBs in the initial microstructure and for the bulk solute concentration of 10~ppm. Dashed lines are determined from the mean-field model considering $E_{avg, m}$ as the representative segregation energy, $E_{rep}$, for each solute. (b) Final grain size ($\overline{R}_f$) for different effective segregation energy distribution in the initial mirostructure.}
    \label{fig:GG_aniso_full}
\end{figure}

\begin{figure}[htb!]
    \centering
    \includegraphics[width=0.7\hsize, keepaspectratio]{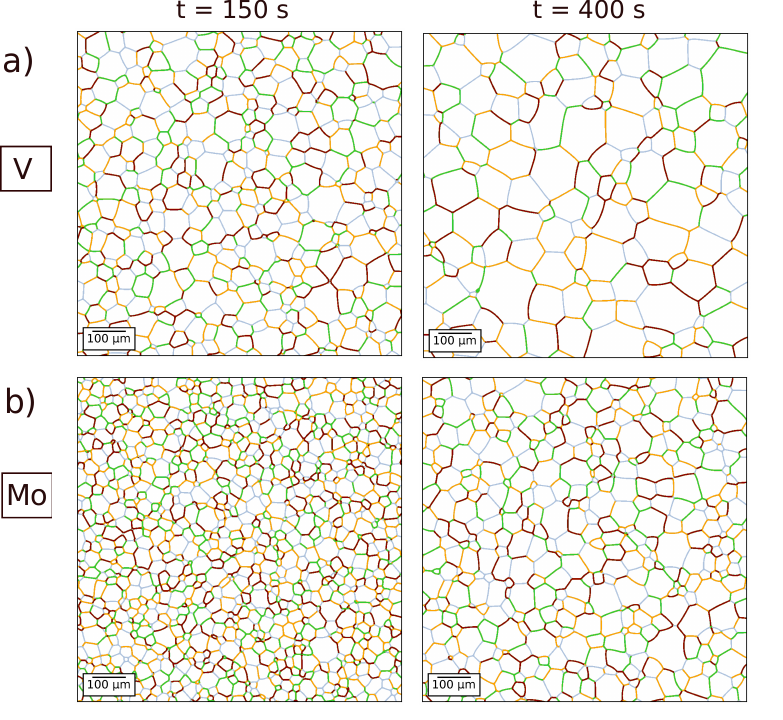}
    \caption{Microstructure evolution for anisotropic grain growth simulations for 10~ppm addition of (a) V (b) Mo after 150~s and 400~s holding time at 1273~K.}
    \label{fig:mic}
\end{figure}

Fig.~\ref{fig:gg_aniso} shows the normalized average grain size evolution from anisotropic grain growth simulations where $\overline{R}_0$ is the initial average grain size. While Nb, Ti, and Mo result in strong grain growth retardation compared to pure Fe, Mn and V have a negligible effect on grain growth rates due to their low effective segregation energies and solute drag pressures. Nevertheless, a parabolic grain growth rate is observed for all cases indicating that all grain boundaries are either in the low-velocity limit or remain almost unaffected by solutes during grain growth.\nopar
 \noindent
{\color{red}{Normal grain growth is observed for each solute, e.g., microstructures for Mo and V addition after 150 and 400 s are shown in Fig.~\ref{fig:mic}, suggesting unimodal grain size distribution that is indicative of normal grain growth in both cases.}} In comparison to the initial microstructure, the proportion of grain boundaries does not change significantly during grain growth, i.e., for the case of Mo, each type of grain boundary is present within 25~$\pm$~3\% even though the number of grains is reduced from $\sim$14000 to $\sim$1200 after 400~s. \
The average grain size evolution in Fig.~\ref{fig:gg_aniso} is used to fit the mean-field model given by Eqn.~\eqref{eqn:ggeq} with $E_{rep}$ as the fitting parameter. $E_{rep}$, for each solute, is found to agree with the $E_{avg, m}$ and the grain size evolution from the mean-field model, i.e., Eqn.~\eqref{eqn:ggeq} with $E_{rep}$=$E_{avg, m}$ is shown with dashed lines in Fig.~\ref{fig:gg_aniso}. 

\sloppy {\color{red} The anisotropic grain growth simulations in the present study, however, represent one of the many possible microstructures where different grain boundary types may not be equally distributed in the microstructure. Thus, to determine the effect of GB distribution on grain growth, we performed three sets of simulations with a single grain boundary type in the microstructure having weak ($\Sigma 41(540)[001]$ GB), moderate ($\Sigma 5(310)[001]$ GB), and strong ($\Sigma 17(410)[001]$ GB) effective segregation energies. In all cases, parabolic grain growth kinetics is observed in agreement with the anisotropic grain growth simulations. The final grain sizes ($\overline{R}_f$) for each case are shown in Fig.~\ref{fig:finalgrainsizes}. It can be observed that the solute trends from anisotropic simulations are consistent with the limiting cases of isotropic simulations indicating that the simulations with an equal fraction of grain boundary types may be considered as a representative case for anisotropic grain growth simulation.} 

\sloppy Further, the role of bulk solute content on grain growth is investigated by determining the effective segregation energy of individual GBs for a given solute concentration that is then used as an input for {\color{red}anisotropic} phase field simulations. For relatively low solute additions, e.g., $c_0$ $<$ 100 ppm, the effective segregation energies are insensitive to the solute content but vary with further increase in solute concentration due to site saturation. As opposed to the atomistic approach of determining the GB excess (Eqn.~\eqref{eqn:DFT_gamma}), the CLS solute drag model does not account for site saturation (Eqn.~\eqref{eqn:CLS_gamma}), and as a result, $E$ effectively changes with the solute concentration. For instance, an increase in Nb content from 100~ppm to 1000~ppm corresponds to a variation in $E$ from -1.11~eV to -0.74~eV for the $\Sigma 5(310)[001]$~GB. 
\begin{figure}[htb!]
    \centering
    \begin{subfigure}{.465\textwidth}
        \centering 
        \includegraphics[width=\hsize]{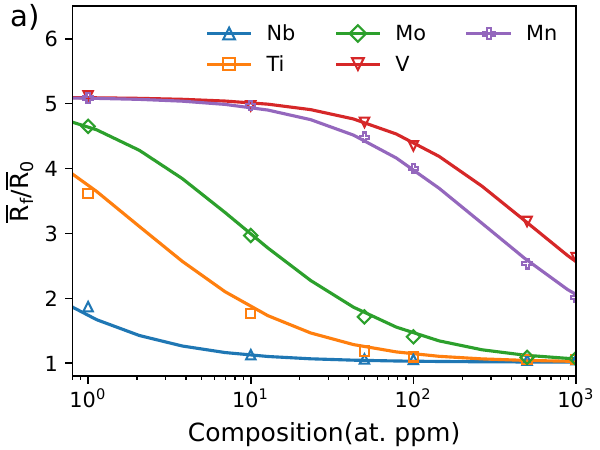}
        \phantomsubcaption
        \label{fig:GG_combined_comp}
    \end{subfigure}%
    \hfill
    \begin{subfigure}{.49\textwidth}
        \centering 
        \includegraphics[width=\hsize]{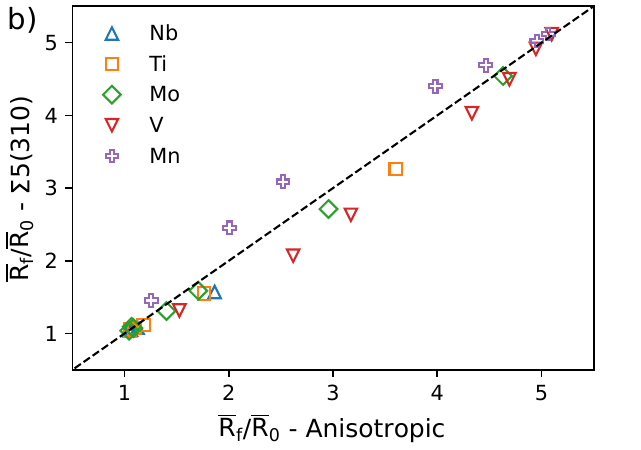}
        \phantomsubcaption
        \label{fig:GG_s5_comp}
    \end{subfigure}%
    \vspace{-7mm}
    \caption{(a) Normalized final average grain size determined from anisotropic phase field simulations as a function of solute concentration. Solid lines are determined from the mean-field model considering $E_{rep}$ = $E_{avg, m}$ (see text). (b) Comparison of normalized final grain size determined from the anisotropic grain growth simulations and isotropic simulations where all grain boundaries are assumed to be equivalent to the $\Sigma5(310)[001]$ tilt grain boundary. }
    \label{fig:GG_comp}
\end{figure}

Fig.~\ref{fig:GG_combined_comp} shows the {\color{red}{final average grain size ($\overline{R}_f$) normalized with initial average grain size ($\overline{R}$)}} obtained from anisotropic simulations that decreases with an increase in bulk solute concentration for all solutes. The solute rankings, however, remain unaffected for a given bulk concentration. The final average grain size determined from the mean-field model considering $E_{rep}$ = $E_{avg, m}$ shows almost perfect agreement with the phase field simulations, further validating the choice of $E_{avg, m}$ for the representative segregation energy. Fig.~\ref{fig:GG_s5_comp} shows a reasonable agreement between the final average grain sizes obtained from isotropic simulations that considered $\Sigma 5(310)[001]$ GB as the representative GB and anisotropic simulations, suggesting that the $\Sigma 5(310)[001]$ GB may be a reasonable choice to determine the solute trends for grain refinement in austenite. It should be highlighted that the trends may be difficult to determine for solutes that have overlapping segregation energies irrespective of their sign, e.g., Mn and V in the present case, which will lead to similar solute drag pressures during grain growth. For instance, isotropic simulations consistently over-predict final grain sizes for Mn and under-predict for V compared to anisotropic simulations. The overall final grain size and solute trends, in these cases, are dependent on the distribution of GB properties in the microstructure. In some steels, however, a higher amount of Mn ($>$0.3 - 2 wt. \%) is added in comparison to V ($<$0.1 wt.\%), and as a result, Mn may become a more relevant element for grain refinement in comparison to V.

\subsection{Solute trend parameter}
The simulations show that the variability in effective segregation energies determined from atomistic calculations has a secondary effect on the solute rankings during grain growth, such that a single representative GB, e.g., the $\Sigma 5(310)[001]$ GB, becomes sufficient to estimate the relative potential of solutes in retarding grain growth in austenite. To further quantify the effects of the bulk solute concentration and effective segregation energy on grain growth, we propose a solute trend parameter given by $|\varGamma_{GB}/D|$, where both the GB excess ($\varGamma_{GB}$) and the bulk solute diffusivity ($D$) can be determined from atomistic simulations and/or experimental measurements.

\begin{figure}[htb]
    \centering
    \begin{subfigure}{.49\textwidth}
        \centering 
        \includegraphics[width=\hsize]{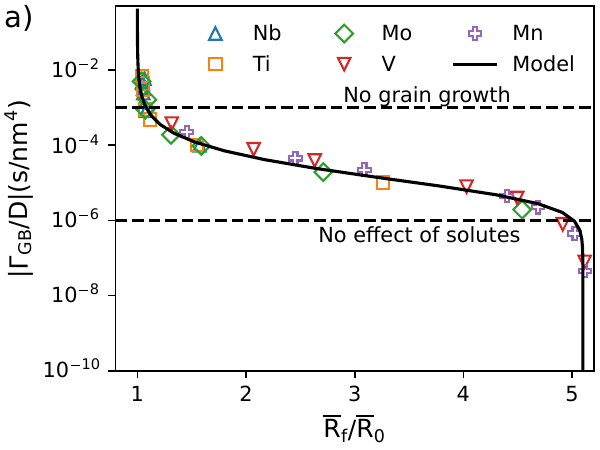}
        \phantomsubcaption
        \label{fig:trendparameter_s5}
    \end{subfigure}%
    \hfill
    \begin{subfigure}{.49\textwidth}
        \centering 
        \includegraphics[width=\hsize]{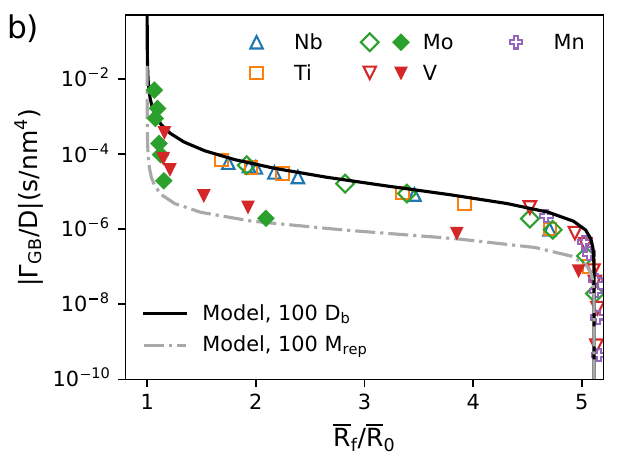}
        \phantomsubcaption
        \label{fig:trendparameter_100D}
    \end{subfigure}%
    \vspace{-7mm}
    \caption{(a) Solute trend parameter ($|\Gamma_{GB}/D|$) as a function of the normalized final grain size determined from the phase field simulations considering isotropic segregation and effective segregation energy corresponding to the $\Sigma5(310)[001]$ tilt grain boundary. Here, $D = D_b$ is used. For different values of the solute trend parameter, Eqn.~\eqref{eqn:ggeq} is used, and the normalized final grain size is shown with a solid line for comparison. (b) Sensitivity of trend parameter with representative GB mobility and trans-GB diffusivity. Solid markers correspond to simulations with increased GB mobility (100 $M_{rep}$), and open markers to simulations with increased trans-GB diffusivity ($D = 100 D_b$), respectively. }
    \label{fig: GG_comp}
\end{figure}

Fig.~\ref{fig:trendparameter_s5} shows the solute trend parameter as a function of the normalized final grain size obtained from the isotropic simulations. Qualitatively, an increase in the trend parameter leads to grain refinement that is in agreement with the conventional solute drag studies where an increase in the GB excess and a decrease in bulk solute diffusivity is correlated with increased solute drag effects~\cite{cahn_impurity-drag_1962}. For different values of the trend parameter, $E$ is determined from Eqn.~\eqref{eqn:CLS_gamma} for a given solute diffusivity and concentration, and the final grain size from the mean-field model is shown with a solid black line. The simulated and predicted final grain sizes for all the solutes fall on the master curve such that the solute additions that correspond to a trend parameter $<$ $10^{-6}$~$\mathrm{s/nm^4}$ have virtually no effect on grain growth, whereas an increase in the trend parameter leads to a transition range, followed by negligible grain growth for trend parameters $>$ $10^{-3}$~$\mathrm{s/nm^4}$. \nopar

The other model parameters, namely the representative mobility, and the trans-GB diffusivity are known with the least certainty. Thus, identical isotropic phase field simulations as above have been conducted, but with trans-GB diffusivities increased by a factor of 100 as well as with a 100 times larger representative mobility for selected cases, i.e., Mo and V. For the latter, the heat treatment time is adjusted according to $M_{rep}$ $\propto 1/t_{sim}$ to obtain 51~$\upmu$m as the final grain size in pure Fe. Fig.~\ref{fig:trendparameter_100D} shows that the transition range is insensitive to the scaling factor in trans-GB diffusion in the former case but shifts by approximately two orders of magnitude to $10^{-7}$-$10^{-5}$~$\mathrm{s/nm^4}$ in the case of simulations with increased GB mobility. In the low-velocity limit, the solute drag pressure from Eqn.~\eqref{eqn:cahnsd} can be approximated by $\alpha c_0 v$, and effective mobility, according to $v = M_{eff}\Delta G$, is determined as $(1/M_{rep} + \alpha c_0)^{-1}$ for a single grain boundary. An increase in the trans-GB diffusivity reduces both $\alpha$ and the solute trend parameter, and therefore, the final grain size shifts along the same master curve. A similar increase in representative mobility, however, has a negligible effect on the effective mobility due to higher contributions from $\alpha c_0$ compared to $1/M_{rep}$, especially for higher solute enrichment. As a result, a nearly identical effective mobility for a given $\alpha c_0$ coupled with smaller heat treatment time results in smaller grain sizes for a higher representative mobility in comparison to simulations with a lower representative mobility. 

\section{Discussion}
\subsection{Comparison with experiments}
\begin{figure}[htb]
    \centering
    \includegraphics[width=0.5\hsize]{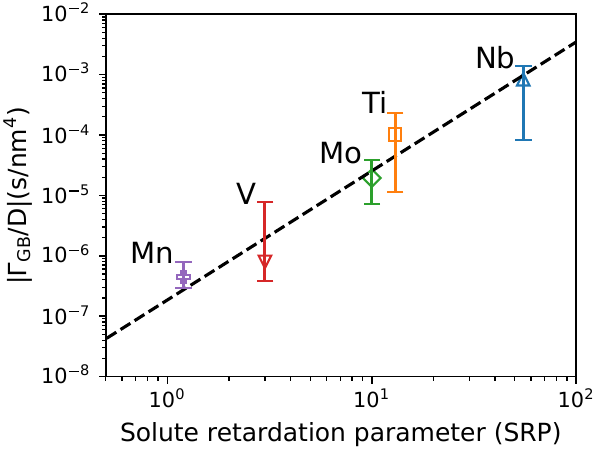}
    \caption{Correlation between the solute retardation parameter (SRP)~\cite{akben_dynamic_1984} and the solute excess normalized by the solute bulk diffusivity (solute trend parameter) considering DFT based segregation energy determined for the $\Sigma 5(310)[001]$ GB and 10~ppm solute additions. The error bars indicate the range of the trend parameter determined for GBs with non-zero GB excess among the nine [001] tilt grain boundaries included in the DFT database. }
    \label{fig:trendparam}
\end{figure}

Relatively little information is available regarding the comparative effects of different solutes on austenite grain growth in ultrapure Fe as opposed to ferrite~\cite{sinclair_effect_2007, hu_time_1970}. Careful investigations of static and dynamic recrystallization in hot-rolled steels have quantified the role of dissolved solutes in reducing the recrystallization rates in austenite. In particular, Jonas and coworkers \cite{akben_dynamic_1981, akben_dynamic_1984, andrade_effect_1983, mcqueen_hot_1995} proposed in a series of studies a solute retardation parameter (SRP) that captures the delay in recrystallization time by the addition of 0.1 at.\% of alloying element in C-Mn steels, which is defined as, 
\begin{equation}
    \mathrm{SRP} = log\left(\frac{t_x}{t_{ref}}\right)\times \frac{0.1}{at. \%} \times 100
\end{equation}
where $t_x$ is the time for the onset of recrystallization, and $t_{ref}$ is the equivalent time in the reference steel (0.06 C-1.4 Mn -0.24 Si-0.024 Al in wt.\%). In these experiments, the authors measured softening kinetics and assumed 10\% softening at 1000~$^\circ{}$C as the time for the onset of recrystallization. Since both recrystallization and grain growth involve migration of high-angle grain boundaries, albeit with different driving pressures, the solute trends are expected to follow a similar relationship even though grain growth rates will be quantitatively more influenced by solute additions due to the smaller associated driving pressures. Note that the SRP was evaluated for one level of solute addition for each alloying element, and as a result, no error bars were reported for the SRP. Further, solutes can have non-linear effects on grain growth rates. Thus, the SRP should be considered as an approximate indicator that characterizes solute effectiveness on grain growth retardation. Fig.~\ref{fig:trendparam} shows the SRP for the investigated alloying elements against the proposed solute trend parameter for 10~ppm solute additions as shown in Fig.~\ref{fig:trendparameter_s5}. The solute trend parameter is also determined for all the grain boundaries shown in Fig.~\ref{fig:e0_compile}, and the range for each solute with finite GB excess is shown as error bars in Fig.~\ref{fig:trendparam}. A correlation between the solute trend parameter and the SRP is evident, indicating that the proposed trend parameter is a good measure to identify solutes that may promote grain refinement in austenite. 

\subsection{Validity of representative grain boundary}
The simulations suggest that the $\Sigma 5(310)[001]$ GB, falling in the midrange of coincident GB energies~\cite{olmsted_survey_2009}, can be considered as a representative grain boundary to identify solutes that may promote grain refinement in austenite. Here, we discuss whether the sample space of investigated GBs is sufficient to support the conclusion of ignoring the structure dependence of the effective segregation energies for solute trend predictions in dilute alloys. Ito and Sawada~\cite{ito_first-principles_2022} demonstrated a monotonic dependence between the Voronoi volume of GB sites and the site-specific binding energies for each solute, indicating elastic contributions as the dominant contributions towards the binding energy. One of the general conclusions is that the GB sites with larger Voronoi volumes show favorable segregation for all the solutes. Each GB has at least one site with a larger Voronoi volume as compared to that of regular lattice sites in the bulk and these larger GB sites control the GB enrichment in dilute systems. The sites investigated in the GB vicinity, i.e., 2.5~$\AA$ from the habit plane, spanned a Voronoi volume from 10 to 14~$\AA^3$ for all the nine grain boundaries such that the largest Voronoi volume in individual GBs is 16-27\% higher than the bulk Voronoi volume in FCC-Fe. Mahmood et al.~\cite{mahmood_atomistic_2022} investigated 22 symmetric tilt GBs and their metastable structures in Al, another FCC system, and observed a similar monotonous change in Voronoi volume and binding energy for Mg segregation. Similar to Ito and Sawada~\cite{ito_first-principles_2022}, they found that the strongest segregating site for Mg is the largest GB site with a Voronoi volume which is 30\% larger than for the bulk sites in Al. Considering Voronoi volume as the primary indicator of segregation energy, it is reasonable to assume that the segregation trends in low coincident “general” grain boundaries will also be qualitatively similar to those observed for high-coincidence GBs.   

\subsection{Solute interactions}
DFT calculations from Ito and Sawada~\cite{ito_first-principles_2022} are applicable to dilute Fe-X alloys with X as the substitutional element, whereas, in reality, carbon in steels can further enhance or hinder solute segregation at austenite grain boundaries. All the investigated solutes in the present study are carbide formers and, therefore, are expected to have a tendency to co-segregate with carbon at the grain boundaries. Using a thermodynamic segregation model, for instance, the interaction coefficient between Mn-C ($\omega_{Mn-C}$) is obtained as -0.5~eV~\cite{enomoto_evaluation_1988}, whereas DFT simulations suggest the interaction coefficient to vary between -0.1~eV to -0.55~eV depending on different GB sites for Mn and C in a $\Sigma 3(111)$ GB in BCC-Fe~\cite{wicaksono_interaction_2017}. In low carbon steels, the carbon concentration profile in the austenite grain boundaries ($c_C$) reaches approximately 2 - 10 at.\%~\cite{enomoto_evaluation_1988, li_segregation_2015}. A first estimation would, therefore, suggest an increase in the effective segregation energy of Mn by $\omega_{Mn-C} c_C$, i.e., -0.01 to -0.05~eV in the presence of carbon. A similar increase in the effective segregation energy is expected for other carbide-forming solutes~\cite{enomoto_evaluation_1988} that will increase the solute trend parameter marginally for these solutes for a given bulk composition. Additionally, the experiments that quantified SRP implicitly included the effect of carbon interactions~\cite{akben_dynamic_1981, akben_dynamic_1984, andrade_effect_1983, mcqueen_hot_1995}, and their agreement with the solute trend parameter further suggests that the solute trends established in the present study will not change significantly by including X-C interactions. On the other hand, solutes such as Si, Ni, and Co show repulsive interactions with carbon such that different solute trends may be obtained in the presence of carbon~\cite{enomoto_evaluation_1988}. As a result, these interactions should be explicitly considered when determining the solute trends for these dissimilar solutes in terms of their interaction with C.

\section{Conclusions}
In this study, we have performed two-dimensional phase field simulations that consider binding energy profiles of several solutes (Nb, Ti, Mo, V, Mn) at different grain boundaries in FCC-Fe to determine the relative solute trends that affect austenite grain growth. The simulations suggest the solute ranking as Nb $>$ Ti $>$ Mo $>$ V $\approx$ Mn in order of their effectiveness to retard grain growth rates. Anisotropic phase field simulations indicate a secondary role of segregation anisotropy on the solute trends such that different combinations of grain boundaries in the initial microstructure lead to identical solute trends. As a result, the $\Sigma 5(310)[001]$ GB, which is easily accessible to DFT simulations, is identified as a representative grain boundary. To further quantify the solute effects, we propose a solute trend parameter defined as the ratio of GB enrichment and trans-GB diffusivity that quantitatively captures the grain size variation in austenite for a wide range of solute additions. The agreement of the solute trend parameter with experimental observations indicates that a representative GB, such as the $\Sigma 5(310)[001]$ GB, may be useful in identifying solute elements that promote grain refinement in austenite using high-throughput atomistic simulations. 
%In this regard, the present approach may further benefit from considering solute interactions with carbon at the grain boundary, especially for solutes that have repulsive interactions with carbon, e.g. Si in steels.  

The presented approach can be used to further explore alloying strategies to control austenite grain size, e.g. during austenite conditioning. The present analysis is based on binary systems where solutes are expected to have similar interactions with C. An extension to multi-component steels and different bulk solute contents is possible within the limit of significantly small interactions between these substitutional solutes. For example, 100~ppm of Nb (Fe-0.017 wt.\% Nb) in solid solution is expected to have a stronger effect on austenite grain size at 1273~K in comparison  with 1000~ppm  of Mn (Fe-0.1 wt.\% Mn), despite  the lower bulk concentration of Nb.
%A comparison of 100 at. ppm of Nb addition (Fe-0.017 wt\% Nb) with 1000 at. ppm. addition of Mn (Fe-0.1 wt\% Mn) suggests a stronger effect of Nb on austenite grain size at 1273 K despite its lower bulk concentration. 
%For such comparisons in metals and alloys, however, special care needs to be taken to use the dissolved solute content instead of the alloy concentration, since the carbide-forming elements may be trapped in precipitates and may remain unavailable for solute segregation and solute drag.

\section*{Acknowledgements}

The authors gratefully acknowledge the financial support under the scope of the COMET program within the K2 Center “Integrated Computational Material, Process and Product Engineering (IC-MPPE)” (Project No 886385). This program is supported by the Austrian Federal Ministries for Climate Action, Environment, Energy, Mobility, Innovation and Technology (BMK) and for Digital and Economic Affairs (BMDW), represented by the Austrian Research Promotion Agency (FFG), and the federal states of Styria, Upper Austria and Tyrol. This research was funded also in part by the Austrian Science Fund (FWF) (P 34179-N).

%% References
\section*{Data Availability}
Data will be made available on request.

%\bibliographystyle{unsrt}
\bibliographystyle{elsarticle-num}
%\bibliography{referencesAyush}
%\bibliography{Ayush}
\bibliography{trendprediction_ayush2}






% \end{thebibliography}

\end{document}